\documentclass{article}

\usepackage{amsfonts,amsmath,amssymb,amstext,latexsym}
\usepackage{balance}
\usepackage{enumerate}
\usepackage{graphicx}
\usepackage{ifpdf}
\usepackage{ifthen}
\usepackage{multirow}
\usepackage{color}
\usepackage[usenames,dvipsnames,table]{xcolor}
\usepackage{xspace}

\newcommand{\eqskip}{2pt}
\newcommand{\thetitle}{Daleel: Simplifying Cloud Instance Selection Using Machine Learning}
\newcommand{\theauthors}{Faiza Samreen, Yehia Elkhatib, Matthew Rowe, Gordon S. Blair}
\newcommand{\thekeywords}{Cloud computing, Machine learning}
\ifpdf
	\usepackage[
		colorlinks=true,%
		linkcolor=black,%
		citecolor=black,%
		filecolor=black,%
		urlcolor=black,%
		pdftex,%
		pdftitle={\thetitle},%
		pdfsubject={Using machine learning to develop an adaptive cloud deployment policy},%
		pdfauthor={\theauthors},%
		pdfkeywords={\thekeywords{}, \theauthors, IaaS selection, cloud migration, Lancaster, NOMS},%
		unicode,bookmarks=false]{hyperref}
\fi

\newcommand*{\eg}{\textit{e.g.}\@\xspace}
\newcommand*{\ie}{\textit{i.e.}\@\xspace}
\makeatletter
\newcommand*{\etc}{%
    \@ifnextchar{.}%
        {etc}%
        {etc.\@\xspace}%
}
\newcommand*{\etal}{%
    \@ifnextchar{.}%
        {et al}%
        {et al.\@\xspace}%
}
\makeatother

\newcommand{\fig}[1]{Fig.~\ref{#1}}

\begin{document}

\title{\thetitle}

\author{
	\theauthors\\
	School of Computing \& Communications, Lancaster University, UK\\
	Email: \{i.lastname\}@lancaster.ac.uk\\
	\textbf{\textcolor{red}{This is a pre-print.}}\\
	\textbf{\textcolor{red}{The final version is available on IEEEXplore}}\\
	\textbf{\textcolor{red}{(ISBN 978-1-5090-0223-8)}}
}

\maketitle

\begin{abstract}
    Decision making in cloud environments is quite challenging due to the diversity in service offerings and pricing models, especially considering that the cloud market is an incredibly fast moving one. In addition, there are no hard and fast rules; each customer has a specific set of constraints (\eg budget) and application requirements (\eg minimum computational resources). Machine learning can help address some of the complicated decisions by carrying out customer-specific analytics to determine the most suitable instance type(s) and the most opportune time for starting or migrating instances. We employ machine learning techniques to develop an adaptive deployment policy, providing an optimal match between the customer demands and the available cloud service offerings. We provide an experimental study based on extensive set of job executions over a major public cloud infrastructure.
\end{abstract}


\section{Introduction}
\begin{figure}[t]
    \centering
    \includegraphics[width=0.8\textwidth, trim=1cm 1cm 1cm 1cm, clip]{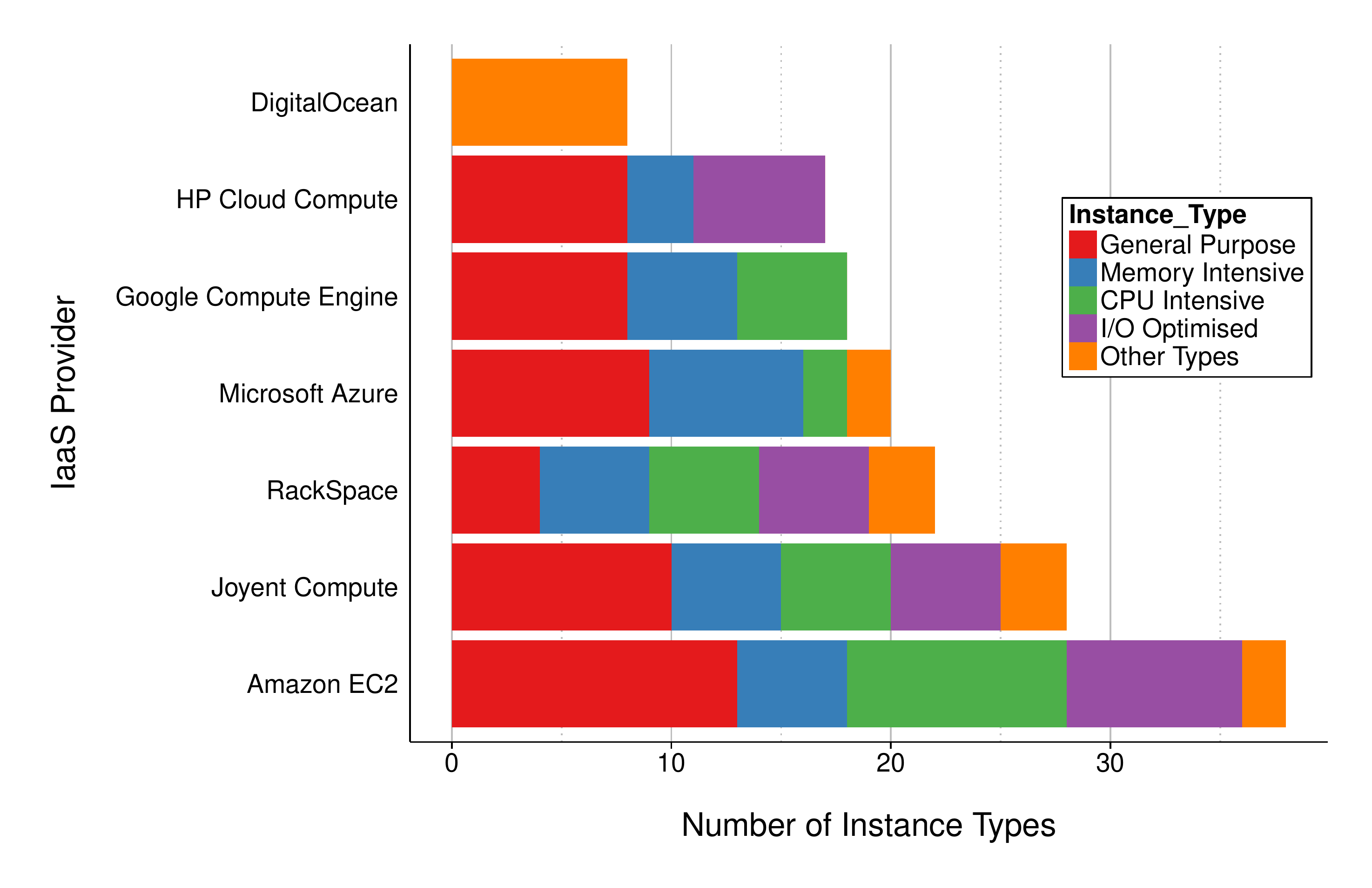}
	\caption{The number of instance types offered by the major IaaS vendors, as of 25\textsuperscript{th} of August 2015.}
	\label{fig:instance-types}
\end{figure}

Users of Infrastructure-as-a-Service (IaaS) provisions are faced with a composite decision:
\begin{itemize}
	\item Which provider should she choose?
	\item What instance type(s) would provide her with the cost:performance ratio that suits her needs? 
	\item Does the time or day at which she requests these resources affect how her application runs?
\end{itemize}
A customer has to choose between dozens of different instance types, as illustrated in Figure~\ref{fig:instance-types}. 
The answers to the above questions are highly subjective; each customer application needs careful consideration of its requirements against the various market offerings.
Further complications are manifested due to the disparate pricing models adopted by different cloud service providers (CSPs) and the rapid evolution of the ecosystem as a result of market forces.
As such, a customer entering the market is overwhelmed with a host of difficult questions without much support for such decision making. 

We argue that we can help answer such questions in a systematic and evidence-based manner.
This is not only to assist customers entering the market, but also to provide guidance to those wishing to migrate deployments between CSPs to enhance Quality of Service (QoS), reduce cost, or to honour other non-functional requirements (\eg legislation, disaster recovery, business continuity).

In this paper we present \emph{Daleel}, a multi-criteria adaptive decision making framework that is developed to find the optimal IaaS deployment strategy. 
We take a first step by focusing on one CSP in order to answer the question: \textit{Which instance type and what time are best for a given customer application?}. 
After gathering substantial profiling evidence, we employ machine learning to gain insight into the expected performance of an application on the calibrated CSP.

Our contributions are as follows:
\begin{itemize}
    \item Daleel, a framework to support adaptive decision making in IaaS environments. We consider two QoS attributes as criteria: instance price and application execution time.
    \item An extensive analysis of variability of Amazon EC2 instances, a leading IaaS CSP. We use more than $5,000$ application runs for this purpose.
    \item Using multivariate polynomial regression (\ie with multiple predictors) to evaluate Daleel's ability to predict application execution time on EC2 configurations.
\end{itemize}


\section{Related Work}
\label{sec:rw}

\subsection{Application Management Frameworks}

A number of \textit{cloud brokering} frameworks have been developed by industry and open source communities to intermediate between cloud customers and providers~\cite{Barker15broker}. These carry out some tasks on behalf of the customer such as arbitrage, aggregation and integration. 
We classify such solutions as either hosted or deployable. \emph{Hosted} services are externally managed by third-party stakeholders and do not provide information about how the application is being provisioned. Examples include RightScale Cloud Portfolio Management, 
enStratus, 
xStream, 
and CliQr. 
In contrast, \emph{deployable} services rely on open source software that could be operated either internally by a corporation or externally as a grey-box service.
Apache Brooklyn, 
Scalr, 
Standing Cloud, 
and Aelous 
are a few examples.

The solutions mentioned thus far tackle interoperability to reduce application deployment friction, but do not support adaptive decision making. This feature is still largely lacking from cloud brokerage solutions~\cite{Samreen2014adaptive}, although some efforts have started to surface, \eg STRATOS~\cite{Pawluk2012stratos}, 
MODACloud~\cite{Gupta2015risk}, 
Cloud4SOA~\cite{Kamateri2013cloud4soa}, 
mOSAIC~\cite{Petcu2013practice}, 
ARTIST~\cite{menychtas2014software}, 
Broker@Cloud~\cite{Patiniotakis2014pulsar}, and 
\cite{Duplyakin2013rebalancing}.  
We conjecture that there is still a long way to go in terms of providing dynamic decision making that can effectively optimise deployment to the specific functional and non-functional requirements on a per-application basis. 
Specifically, which resource type is most cost effective while considering the good performance for my application? 
This is the potential domain for machine learning that can contribute for enhancement of multi-cloud management by taking appropriate decisions to cater the application and application owner requirement.

\subsection{Machine Learning}

Machine learning can contribute immensely by taking appropriate decisions to cater to specific application requirements.
Machine Learning has proved its potential for producing prediction and optimisation solutions in various fields. 
It has been applied in cloud computing towards resource scaling~\cite{Kupferman2009scaling}, forecasting~\cite{Caron2010forecasting}, and dynamic resource provisioning~\cite{Islam2012empirical,Bankole2013predicting,Gao2013mldco}.
We aim to apply a similar methodology but for the benefit of customers selecting between IaaS resources.

\section{Daleel}
\label{sec:admb}
\emph{Daleel} (meaning `guide' in Urdu) is a multi-criteria adaptive decision making framework. 
It equips a cloud customer with evidence-based knowledge of the IaaS setup specification that is optimal for their particular application.

\subsection{Architecture}

\begin{figure}[th]
    \centering
    \includegraphics[width=0.9\columnwidth, trim=0.5cm 0.65cm 0.65cm 0.65cm, clip]{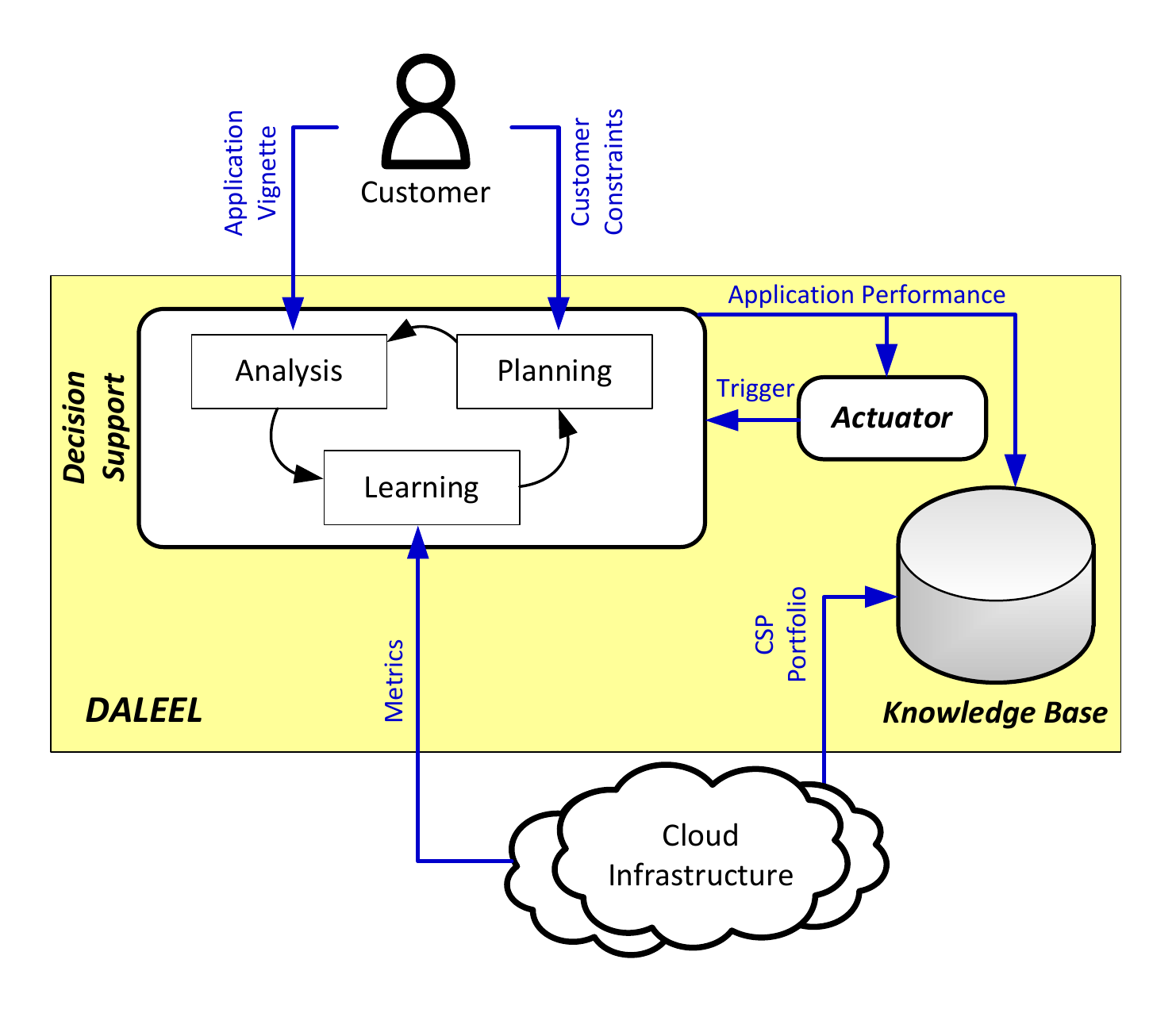}
	\caption{The Daleel Architecture.}
	\label{fig:arch}
\end{figure}

Daleel's architecture (depicted in \fig{fig:arch}) consists of three main modules: \emph{Decision Support}, \emph{Actuator}, and \emph{Knowledge Base}.
At the heart of the architecture is the \textbf{Decision Support} module which relies on a three phase process that continuously operates throughout the application life cycle to predict application performance. These phases are: \emph{Analysis}, \emph{Learning}, and \emph{Planning}.
They carry out different but complimentary operations to acquire deep knowledge of the available cloud deployment options and how suitable they are to a given application.
The \textbf{Actuator} triggers the Decision Support module into operation based on application performance.
The \textbf{Knowledge Base} holds data collected by the Decision Support module as well as CSP portfolios. The latter contains data obtained through APIs and web scraping on CSP resource provisioning levels, resource metadata, and pricing models.

The customer provides an application vignette which is a high level description of the application computational requirements in the form of a short set of key-value pairs.
The customer also indicates their constraints 
such as minimum QoS, availability, location, and budget.

We now describe how the three Decision Support phases and the Actuator module work.

\subsection{Analysis Phase}
The first stage comprises of a profiling procedure that is based on time series analysis. Application profiling is an effective way of tracking application behaviour under different deployment setups. This can be carried out \textit{live} on shared cloud infrastructures (whether public or private), or \textit{offline} in a completely controlled and isolated virtual environment.
The obtained traces record different metrics such as CPU and memory utilisation, paging and caching information, \etc
Together these constitute the application profile that can be used to predict deployment options that can suit the application and customer requirements. 

Aggregating different application profiles builds up the Knowledge Base with information about application descriptions and their behaviour on different deployment setups. This is used to infer performance of a not-yet-profiled application based on its vignette (\ie general description).

\subsection{Learning Phase}

This second phase comprises of a learning procedure that receives the following information from the Analysis phase: performance traces, cloud resources portfolio, and the application vignette. A critical task is to derive a prediction model for the cloud provider in question, which shows the QoS variations and offered services in order to help the following phase (\ie Planning) in achieving optimal deployment that caters to the customer constraints. 

The Learning phase aims to learn the prediction model to accurately predict the cost of application execution in terms of performance and virtual machine (VM) price. It also aims to achieve a better understanding of the correlation between the predictors and the response in order to infer some relationship for future prediction. 
These are quite difficult aims for which different machine learning techniques are explored as no one technique is considered to be the best for all data sets. 
We employed different regression methods as a prediction function. The response variable in our case study is application execution time that involves a continuous quantitative output value. 
This is often referred to as a \textit{regression problem}. 

\subsection{Planning Phase}

The third phase takes input from the Learning phase in the form of a prediction model which can generate a vector output based on the input requirements of the customer. 
The Planning logic is designed to support a \emph{multi-criteria decision making} problem where a set of vectors describing the performance is the Learning outcome. For the purposes of this study, we are targeting two QoS attributes as our criteria, namely VM price and application execution time. 
Various methods are being used by multi-criteria decision making such as weighted sum~\cite{Marler2012weightedsum}, weighted product~\cite{Mateo2012weightedproduct}, VIKOR~\cite{Liao2013vikor}, and PROMETHEE~\cite{Brans2005multicriteria} (see~\cite{Samreen2014adaptive} for more details).
We intend for our decision making support to include such multi-criteria techniques while considering more than two QoS attributes (\ie more than one customer-dictated objective).

\subsection{The Actuator}

The Actuator triggers the Decision Support module into operation at different times.
This could be based on thresholds set according to the customer constraints on application QoS, application load, or Knowledge Base information (\eg change in a provider's portfolio).
Such triggers will launch new Analysis and Learning cycles, or will activate the Planning logic to begin migration to a new cloud infrastructure. 
Migration between different cloud infrastructures is a big challenge in its own right and is outside the boundaries of this work. However, the Planning logic could easily be extended to incorporate migration methods, \eg~\cite{hadley2015multibox}.

\section{Analysis of Variability in IaaS Offerings}
\label{sec:anlyss}
Selecting specifications of a cloud-based infrastructure is not an easy or straight-forward task, especially due to the fact that there is considerable amount of performance variability at any service provisioning tier.
Our initial step is to gather enough information to analyse such variability.
We achieve this through extensive experiments over the Amazon Elastic Cloud Compute (EC2) IaaS offerings. In this section we explain the experimental setup, profiling procedure, and performance variability analysis. \S\ref{sec:eval} will detail model development and learning evaluation based on the profiled data.

\subsection{Methodology}
The overall objective of conducting this evaluation is to find the performance variations on different node configurations at different times of the day. This experiment is conducted on EC2, the leading IaaS provider with a 57\% market share~\cite{sotc2015}. 
We run over different instance types and throughout the seven days of the week to investigate temporal variations.

\subsubsection{Infrastructure}
All instances used were 64-bit Ubuntu Linux of different capacities as detailed in Table \ref{tab:instance-types}.
Note that `vCPU' indicates the number of virtual cores assigned to a VM.
An `ECU' refers to an \textit{EC2 Compute Unit}; Amazon does not advise about how an ECU relates to physical processing speed; it only assures that it is a standard unit across its IaaS offerings\footnote{\url{http://aws.amazon.com/ec2/faqs/}}.
`Price' refers to the hourly charge for running a VM of the referenced instance type.

\begin{table}[ht]
    \def\arraystretch{1.2}
    \setlength{\tabcolsep}{4pt}
    \centering
    \caption{The computational specification of EC2 instances.}
    \begin{tabular}{ccccccc}
    \hline
    \textbf{Series} & \textbf{Node} & \textbf{vCPU} & \textbf{ECU} & \textbf{RAM} & \textbf{Storage} & \textbf{Price}\\
    &&&&(GB)&(GB)& (\$/h)\\
    \hline
    T2 (General & t2.small & 1 & Var. & 2 & 20 & 0.026\\
    \cline{2-7}
    Purpose) & t2.medium & 2 & Var. & 4 & 20 & 0.052\\
    \hline
    M3 (General & m3.medium &1 & 3 & 3.75 & 4(S) & 0.070\\
    \cline{2-7}
    Purpose) & m3.large & 2 & 6.5 & 7.5 & 32(S) & 0.140\\
    \hline
    C4 (Compute & c4.large & 2 & 8 & 3.75 & 20 & 0.116\\
    \cline{2-7}
    Optimised) & c4.xlarge & 4 & 16 & 7.5 & 20 & 0.232\\
    \hline
    \end{tabular}
    \label{tab:instance-types}
\end{table}

Amazon provides differentiated series of instance types, catering to various application needs (\eg compute-intensive, memory intensive, I/O-intensive, \etc). Each series contains a number of instance types with different setups of computational resources.
We targeted the General Purpose series T2 and M3 as well as the Compute Optimised series C4 in order to evaluate varying combinations of resource capacity over a relatively wide price range. 
Only on-demand instances were used for this experiment. These have no long term commitment and are charged on a pay-as-you-go basis at an hourly rate. All instances were chosen to be located in the {eu-west-1} availability zone, hosted in Ireland.

We are not aware of how EC2 virtual cores are pinned to physical cores. Amazon EC2 uses the Xen hypervisor to host the VM instances but do not provide the details of scheduling algorithms used by the hypervisor. From running our experiments, we could not find any firm details for parallel workload and so are not aware of the interference effects. This, however, is not our focus.

\subsubsection{Application \& Execution}
\label{sec:anlyss:app}
Our use case application was VARD~\cite{baron2008vard2}, a tool designed to detect and tag spelling variations in historical texts, particularly in Early Modern English. The output is aimed to improve the accuracy of other corpus analysis solutions.
VARD is a single threaded application that is highly memory intensive.
It holds in memory a representation of the full text, as well as various dictionaries that are used for normalising spelling variations.
Experiments were continuously repeated using a fixed set of input texts over a period of seven days with a delay of ten minutes between each pair of runs.
The Linux tools \texttt{vmstat}, \texttt{glances} and \texttt{sysstat} were used to continuously monitor resource utilisation.

\subsection{Variation Due to Instance Type}

\begin{figure}[th]
    \centering
    \includegraphics[width=0.8\textwidth, trim=0cm 0cm 0cm 1cm, clip]{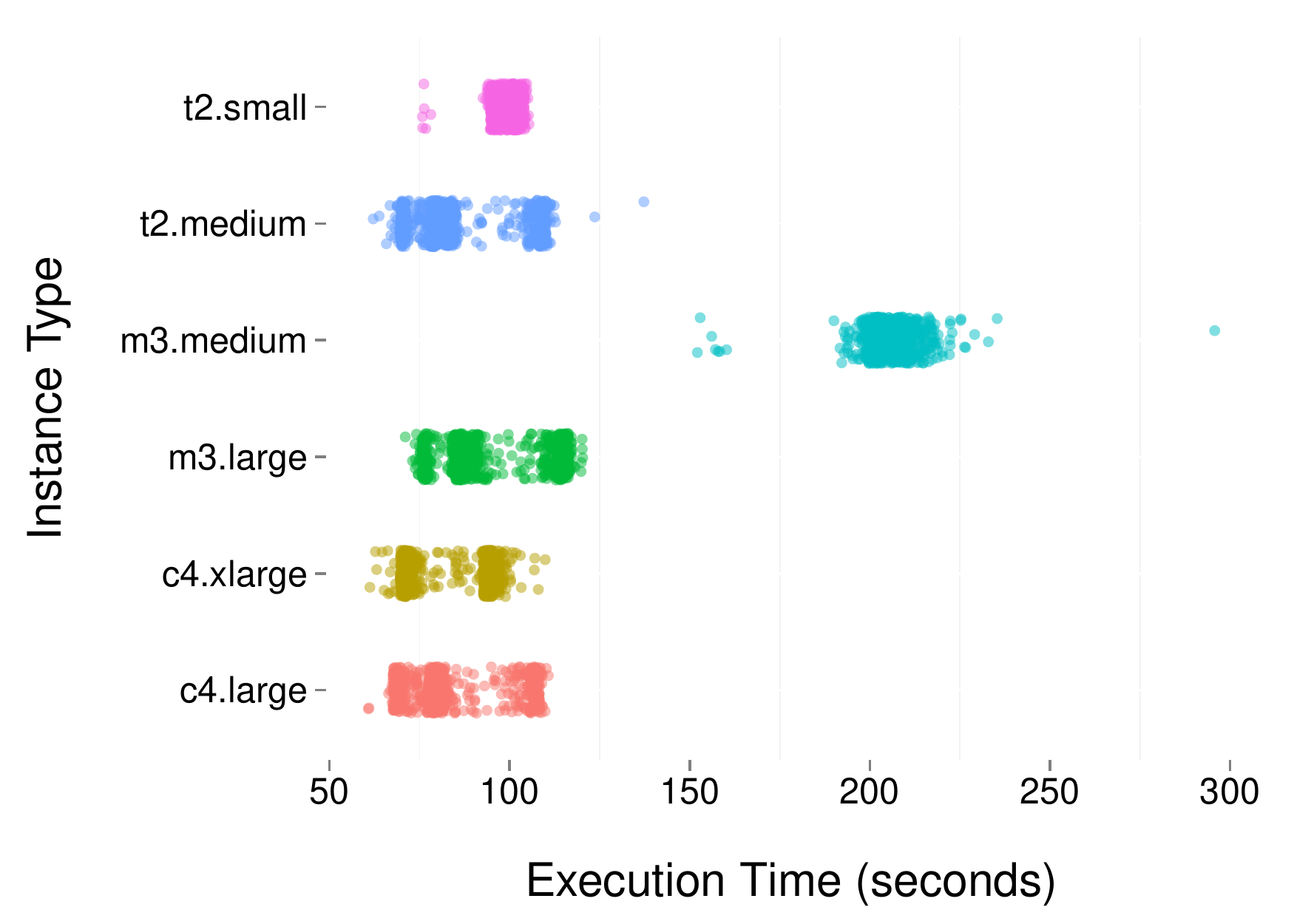}
	\caption{Application execution time over different cloud instance types.}
	\label{fig:execution-time}
\end{figure}

We investigate the performance of running VARD on VMs of different instance types. The results are summarised in \fig{fig:execution-time} where every dot represents the execution time of one run. 
Shorter execution times reflect a lower hourly rate over a full workload. 
There are several striking observations.

First, contrary to intuition, m3.medium (a memory-rich instance) is of consistently poor performance.
We also observe that c4.large surpasses both m3.medium and m3.large in performance.
In fact it is on par with c4.xlarge, which is twice both in specification and cost.

Overall, the T2 series offers by far the best value for money.
A possible explanation is the \textit{CPU Credits} scheme, offered only on the T2 series, which enables customers to collect credits for idle instances and later spend them when full CPU utilisation is needed. 
T2 instances are thus good for applications that do not consistently fully use the CPU, accumulating CPU credits at a steady rate.
but it also means that there is a degree of uncertainty associated with an application's performance that depends on its idle time.

\subsection{Variation Due to Time}

\begin{figure}[htb!]
  \centering
   \includegraphics[width=\textwidth, trim=0cm 0.25cm 0cm 0.5cm, clip]{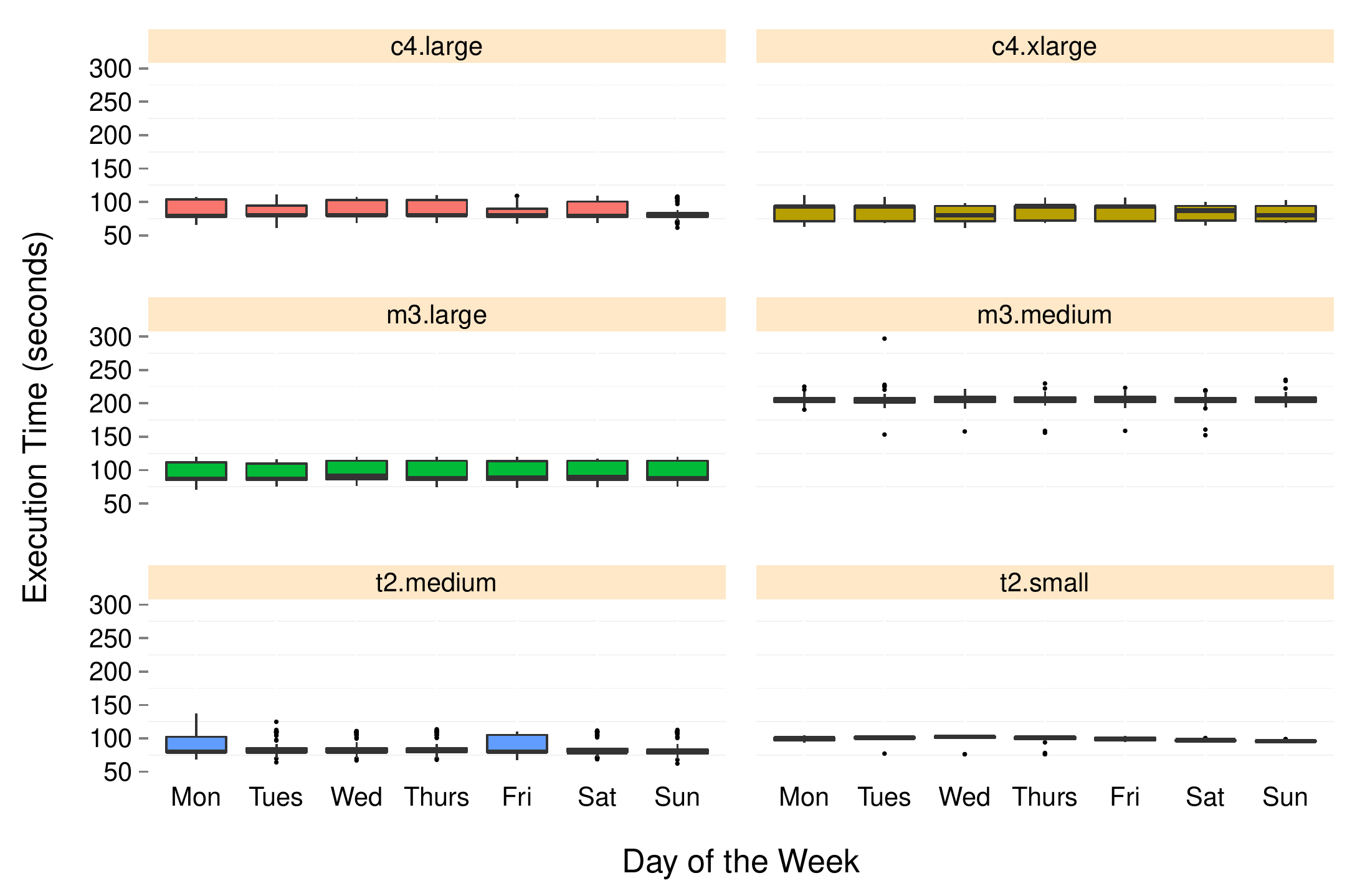}
\caption{Dispersion of application execution time during all days of the week on different EC2 instance types. Notice that all graphs have the same y-axis range apart from m3.medium.}
\label{fig:dayvariance}
\end{figure}

We now turn our attention to uncertainty in application performance due to the time at which they are executed. This is depicted by the box-plots in \fig{fig:dayvariance}.

The T2 series offers the least RAM but exhibits the least variance in performance between the different days of the week. 
m3.medium VMs display application execution times that are fairly high albeit predictable: the median and quartiles show very little variation across the days of the week. 
m3.large also offers quite predictable performance across the week, with a narrow first quartile which is favourable.

The two C4 instance types portray contrasting performances. 
c4.large is rather predictable with a steady median and right skewness (\ie a very narrow first quartile). On the other hand, dispersion in the c4.xlarge instances is more towards the high end of application execution time with a median that is less regular: less left skewness is observed on Wednesday, Saturday and Sunday.

This could be down to different reasons such as demand from other users, the provider's resource sharing algorithms, and the provider's energy efficiency policy.
These are difficult attributes for us to ascertain from the outside.
Nevertheless, we detect certain regularities that helps us determine the predictability of application performance at different time.

\subsection{Lessons Learned}

We investigated how variable the performance obtained from different IaaS settings could be, making the execution of a simple application rather uncertain. 
This demonstrates that public IaaS offerings are to a great extent black boxes. First, selecting instance types solely based on their advertised resource specifications is not necessarily optimal.
Second, selecting which day of the week to run an application could result in significant variation in performance.

\section{Learning Evaluation}
\label{sec:eval}
The confirmed performance variability serves as our motivation to equip users with some certainty when consuming IaaS resources.
We apply different machine learning algorithms to be able to predict the best IaaS deployment setup for a certain application. We again use the VARD application (see Section~\ref{sec:anlyss:app}) as a use case, with a goal to predict the optimal resources and most opportune time for starting an EC2 instance to execute VARD. 
We first describe how we look for and assess the best models, then we detail the outcomes of our learning investigation.

\subsection{Model Development and Evaluation Method}
\label{sec:eval:meth}
The core technique of our methodology that can effectively predict execution time is based on polynomial regression. Polynomial regression is an approach of non-linear fit to data~\cite{hastie2013elements}. It extends the linear model by adding additional predictors that are obtained by raising each of the original predictors to a power. 
We take application execution time as a response variable, whilst a list of other variables as candidate predictors: RAM, vCPU, processor speed, hypervisor, storage, day, time, application input parameters, \etc

Considering both prediction and inference based learning techniques, we follow the procedure outlined below in order to get a robust model that can accurately predict the response using the predictors.
		\begin{enumerate}
			\item Split the data into two sets: a training set to be used for learning, and a test sample for assessment and model evaluation. For current evaluation we split the data set into training and test set with a ratio of 57\% and 43\% respectively.
			\item Train the model on the training set.
			\item Assess the accuracy of the model using resampling (\eg cross validation and bootstrapping), on the training set. Resampling methods repeatedly draw samples and refit the model on each sample to get additional information about the fitted model's performance such as variability estimates of regression fit. Cross validation is one of the widely used resampling methods for model selection. 
			We used k-fold cross validation computed by averaging the Mean Squared Error (MSE) for k-folds over the training sample.
            The MSE serves as a risk function for an estimator to measure 
            the difference between the estimator and estimated value. 
			\item Check goodness of model fit using statistical tests like p-value, $R^2$, RSE, and F-statistics. $R^2$ measures the proportion of variability in the response variable that is explainable by the predictors. The Residual Standard Error (RSE) shows the actual deviation of the response from predicted, and measures the lack of fit for a model. F-statistics (also referred to as \textit{fixation indices}) is a measure to reject a null hypothesis and to show the overall significance of a model. 
		\end{enumerate}

\subsubsection{Polynomial Fit}

The multivariate polynomial model is a special case of a basis function approach that we used in our learning model. 
The idea of using a basis function is to have a transformation that can be applied to a variable $X$: 
$b_1(X), b_2(X), \dots b_k(X)$. 

Basis functions are fixed and known, 
hence the least square approach can be used to estimate the unknown regression coefficients in the model above. A robust polynomial model is build using profiling results. This model is an attempt to predict execution time using two significant predictors RAM and vCPU as well as an additional one: day of the week. This third predictor cannot describe the underlying distribution function on its own; instead it presents a meaningful outcome in a combinatorial way. 
This polynomial regression based formula takes the following form: 
\begin{multline*}
F(x) = \beta_{01} + \beta_{11}x_1 + \beta_{21}x_1^2 + \\
  \beta_{02} + \beta_{12}x_2 + \beta_{22}x_2^2 + \beta_{32}x_2^3 + \\
  \beta_{03} + \beta_{13}x_3 + \beta_{23}x_3^2 + \beta_{33}x_3^3
\end{multline*}

This model is considered a successful attempt towards prediction at a fine grained level. 
It has the lowest MSE compared to other models evaluated in next section. The planning phase takes this model as an input along with substantial details of customer constrains and outputs the suitable configuration based on the metric calculated by the planner.

\subsection{Model Accuracy Analysis}
To evaluate the accuracy of our model we compared it with different learning models using the standard methods described in subsection~\ref{sec:eval:meth}. 
Due to the lack of previous models, we used other learning techniques as baseline for comparison, namely linear regression, ridge regression and Lasso. The same dataset and methodology were used to extract and evaluate the results. 

\subsubsection{Baseline Models}
Linear models are relatively simple to implement and can provide good interpretation and inference. For accurate coefficient estimates, it uses the least square criteria~\cite{Mitchell1997ml}. 

In extension to our assumption about linear regression we tried to fit the model containing all variables using a technique that shrinks the coefficient estimates towards zero. We used the two well-known regularisation techniques for shrinking regression coefficients: ridge regression (also known as \textit{Tikhonov regularisation}~\cite{tikhonov1977ridge}) and Least Absolute Shrinkage and Selection Operator (Lasso)~\cite{Tibshirani1996lasso}.

Ridge regression is similar to least squares but minimises the coefficient estimates with a slightly different quantity of the tuning parameter $\lambda$~\cite{James2014ISL}. 
When $\lambda=0$, the penalty term has no effect and estimates are least square. As $\lambda \rightarrow \infty$ the shrinkage penalty grows and the coefficient estimates approaches zero.

Ridge regression includes all the variables as P predictors in the final model. Highest value of $\lambda$ can reduce the coefficient value but cannot exclude any variable from the resulting model. On the other hand, Lasso overcomes this disadvantage by forcing some of the coefficient estimates to be equal to zero especially when the $\lambda$ value is large enough~\cite{James2014ISL}.

In statistical terms, Lasso uses an $l_1$ penalty while ridge uses an $l_2$ penalty. 
For both these model fits, we chose a range of $\lambda$ values from $\lambda=10^{10}$ to $\lambda=10^{-2}$ in order to evaluate all scenarios starting from the null hypothesis (that contains only the intercept term) to the least square fit, respectively. 

\subsubsection{Diagnostic Assessment}

We now asses the accuracy of the models, as summarised in Table~\ref{tab:training}.

The norm values assessment for Lasso and ridge regression models indicate that none of the $\lambda$ values reduced the MSE. In fact, the best $\lambda$ values (\ie the ones that have minimum MSE, namely $\lambda=2.30$ for ridge and $\lambda=0.03$ for Lasso) have even higher MSE than that when the function is derived to the least square fit. The best $\lambda$ value was figured out using cross validation technique.

Moving on to the other regression diagnostics (not suitable for ridge or Lasso), the $R^2$ statistic provides the proportion of variance explained using the predictor $X$ and so it always takes a value between 0 and 1. The low $R^2$ value for linear regression indicates that this model did not explain much of the variability in the response; much less than half of it, in fact. On the other hand, the polynomial model captures more than 93\% of data variability in terms of response prediction. 

The high F-statistics value for the polynomial model indicates the significance of selected predictors and their relationship with the response variable. The validation set error rate is usually assessed using MSE especially in the case of quantitative response. The MSE values for ridge and Lasso are higher than that of the linear model. However, the same validation set MSE for polynomial fit is considerably smaller than the linear model.
As with $R^2$, we observe a gross reduction for RSE in polynomial fit that estimates the standard deviation of error term which means there is less deviation of predicted response from the true regression line. 

\def\arraystretch{1.2}
\def\colwidth{0.125\columnwidth}
\newcolumntype{P}[1]{>{\raggedleft\arraybackslash}p{#1}}

\begin{table}[th]
    \centering
    \caption{Model assessment over the training dataset.}
    \rowcolors{2}{blue!15}{}
		\begin{tabular}{lP{\colwidth}P{\colwidth}P{\colwidth}P{\colwidth}}
 \hline
\multirow{2}{*}{Diagnostic} & \multicolumn{4}{c}{Model} \\
                            & Linear    & Ridge     & Lasso     & Polynomial \\
 \hline
 MSE (10-fold CV)           & 1159.00   & 2312.69   & 2476.65   & 131.27 \\
 {$R^2$}				    & 0.3741    & --        & --        & 0.9307 \\ 
 F-Statistics 			    & 298       & --        & --        & 5024   \\ 
 RSE                        & 33.77     & --        & --        & 11.55 \\
 \hline
 	\end{tabular}
	\label{tab:training}
\end{table}

\begin{figure}[tbh]
    \centering
    \includegraphics[width=\columnwidth, trim=0cm 0.75cm 0cm 0.75cm, clip]{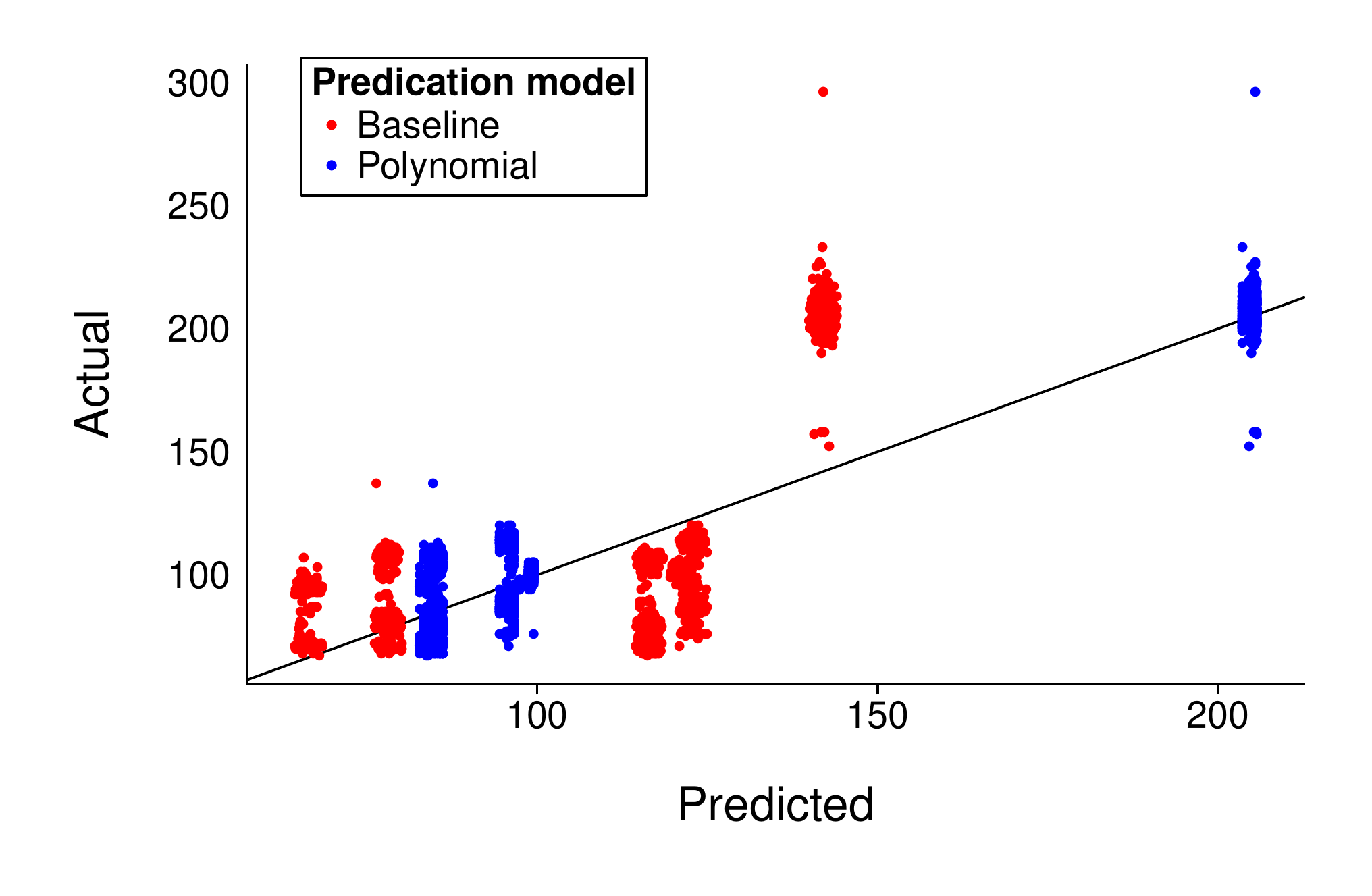}
	\caption{Predicted vs Actual values using the polynomial and linear baseline model over the test dataset. Values closer to the identity line indicate better prediction performance.}
	\label{fig:scatterplot}
\end{figure}

Furthermore, we can check the model visually by plotting the actual response from the test dataset against the predicted. If the model describes the structure of the data appropriately, then the estimated regression curve should be aligned with the data. This visualisation is shown in \fig{fig:scatterplot}. The red points denote the linear model fit, depicting considerably high deviation from the identity line. Ridge and Lasso display similar qualities and are not plotted for clarity. In contrast, the polynomial fit (blue points) is closely aligned to the identity line, proving a far superior prediction capability compared to the linear and other baseline models.

Finally, we assess the models on the test set (43\% of the full data). The results (Table~\ref{tab:test}) confirm the above findings.

\begin{table}[th]
    \centering
    \caption{Model assessment over the test dataset.}
    \rowcolors{2}{blue!15}{}
		\begin{tabular}{lP{\colwidth}P{\colwidth}P{\colwidth}P{\colwidth}}
 \hline
\multirow{2}{*}{Diagnostic} & \multicolumn{4}{c}{Model} \\
                            & Linear    & Ridge     & Lasso     & Polynomial \\
 \hline
 MSE           & 1183.00   & 1185.82   & 1162.80   & 129.84 \\
 \hline
 	\end{tabular}
	\label{tab:test}
\end{table}

\subsection{Model Fitting Outcomes}

\begin{figure}[bt!]
   \centering
   \includegraphics[width=\columnwidth, clip, trim=10cm 8cm 0cm 1.1cm]{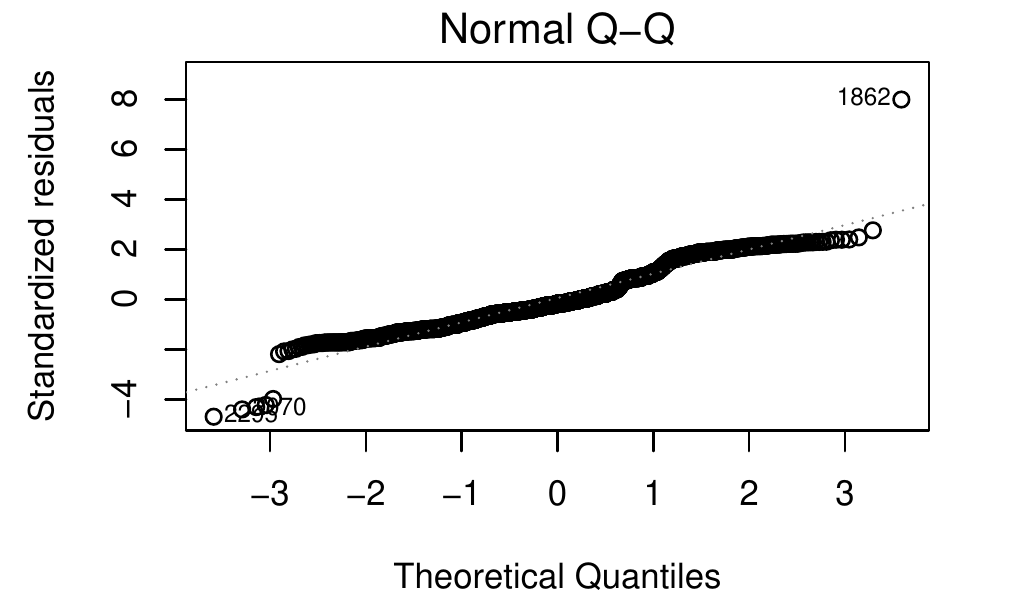} 
   \caption{Q-Q plot of the polynomial model.}
 \label{fig:qqplot}
 \end{figure}

The polynomial transformation has proved to be the best fit to the EC2 data as evaluated using different diagnostic methods and plots.
We also used residual plots to check for possible violations of our assumptions such as non-constant variance and non-normal distribution. If the two distributions are similar, then they would show a constant variance with normally distributed data.
We employed Q-Q plots for this purpose (omitted for space), and they confirmed that the data follows a normal distribution, validating our assumption, and that non-linear transformation works well for our model fit.
In the Q-Q plots of our polynomial model shown in \fig{fig:qqplot}, the points lie on the identity line indicating that the data indeed follows a normal distribution, validating our assumption. This also confirms that the non-linear transformation works well for our model fit.

An interesting finding is that the inclusion of an additional predictor (day of the week) has little but not significant improvement as reflected by $R^2$ in the polynomial model. Nonetheless, it does allow us to be able to predict the optimal deployment time at a day granularity level which is a good contribution to support decision making. 

In summary, we explored the possibility of fitting the data with both linear and non-linear models.
We found a non-linear transformation of the predictors is more suitable due to the non-linear association of data.
Model assessment was done through cross-validation using the MSE which estimates the test errors associated with the learning method to evaluate its performance. 
We applied regression diagnostics to check the assumptions for linear regression with non-linear transformation of the predictors. Non-linear multivariate polynomial model outperformed the linear, ridge and Lasso models as indicated by different factors.

\section{Conclusion}
\label{sec:conc}

Customers are faced with a myriad of choices for deploying applications in the cloud.
Supporting the customer decision making is an under-researched area. Furthermore, there is little understanding of how to implement adaptive decision making that can react to changes in context.

In this paper, we explored the role of machine learning to provide such adaptive decision making. An overall architecture, called \emph{Daleel}, was proposed (\S\ref{sec:admb}) and an empirical evaluation carried out to investigate how to support the key phases of analysis and learning in preparation for the subsequent planning. The analysis focuses on decision making around executing a particular application in a given cloud provider. Specifically, our empirical study focused on decision making around the selection of instance types for the execution of VARD, an application from the field of computational linguistics. The results show that:
\begin{enumerate}[a.]
	\item The performance of instance types can vary significantly over time and is often counter-intuitive (\S\ref{sec:anlyss}).
	\item Machine learning can be highly effective in making performance predictions but care is required to select the right approach (\S\ref{sec:eval}).
\end{enumerate}
 
Looking at the latter aspect in more detail, we explored the possibility of fitting the data with both linear and non-linear models. We found that a non-linear transformation of the predictors to be much more suitable.
This is due to the inherent non-linear association of the data; we found, through extensive experimentation, significant skewness in the performance of different Amazon EC2 instances.

The results from this initial study are encouraging and we are convinced that machine earning has a central role to play in optimising cloud deployments. 
Our first avenue of future work is to consider other application types, such as memory-intensive, processor-intensive, data-intensive and combinations thereof. The Knowledge Base in our Daleel architecture will be used for further investigation of application behaviour and resource utilisation patterns in order to predict based on different clusters of application types.
Second, we plan to incorporate enhanced decision making methods, including consideration of multi-criteria decision making.
Finally, extending the work to consider the management of cross-cloud environments including consideration of policies such as cloud-bursting in hybrid cloud infrastructures~\cite{elkhatib2013experiences}. Cloud brokerage is the key application area here, but it still is in its infancy especially in terms of practical experience~\cite{elkhatib2015building}.

%

\balance{
	\bibliographystyle{abbrv}
	\bibliography{IEEEabrv,daleel}
}

\end{document}